\documentclass[11pt,twoside]{article}

\usepackage{asp2004}
\usepackage{epsf}
\usepackage{psfig}
\usepackage{lscape}

\begin{document}

\title{A Comparison of SDSS Standard Star Catalog for Stripe 82 with Stetson's Photometric Standards     }   

\author{
\v{Z}. Ivezi\'{c}$^{1}$,
J.A. Smith$^{2}$,
G. Miknaitis$^{3}$,
H. Lin$^{3}$,
D. Tucker$^{3}$,
R. Lupton$^{4}$,
G. Knapp$^{4}$,
J. Gunn$^{4}$,
M. Strauss$^{4}$,
J. Holtzman$^{5}$,
S. Kent$^{3}$,
B. Yanny$^{3}$,
D. Schlegel$^{6}$,
D. Finkbeiner$^{7}$,
N. Padmanabhan$^{6}$,
C. Rockosi$^{8}$,
M. Juri\'{c}$^{4}$,
N. Bond$^{4}$,
B. Lee$^{6}$,
S. Jester$^{9}$,
H. Harris$^{10}$,
P. Harding$^{11}$,
J. Brinkmann$^{12}$,
D. York$^{13}$,
for the SDSS Collaboration
}

\affil{
$^1$
Department of Astronomy, University of Washington, Seattle, WA 98115,
$^2$
Department of Physics \& Astronomy, Austin Peay State University,
Clarksville, TN 37044,
$^3$
Fermi National Accelerator Laboratory, P.O. Box 500, Batavia, IL 60510,
$^4$
Princeton University Observatory, Princeton, NJ 08544,
$^5$
New Mexico State University, Box 30001, 1320 Frenger St., Las Cruces, NM 88003
$^6$
Lawrence Berkeley National Laboratory, MS 50R5032, Berkeley, CA, 94720,
$^7$
Department of Astronomy, Harvard University, 60 Garden St., Cambridge, MA 02138
$^8$
University of California--Santa Cruz, 1156 High St., Santa Cruz, CA 95060,
$^{9}$
School of Physics and Astronomy, University of Southampton, Highfield, Southampton, SO17 1BJ, UK, 
$^{10}$
U.S. Naval Observatory, Flagstaff Station, P.O. Box 1149, Flagstaff, AZ 86002,
$^{11}$
Department of Astronomy, Case Western Reserve University, Cleveland, Ohio 44106 
$^{12}$
Apache Point Observatory, 2001 Apache Point Road, P.O. Box 59, Sunspot, NM 88349-0059
$^{13}$
University of Chicago, Astronomy \& Astrophysics Center, 5640 S. Ellis Ave., Chicago, IL 60637
}

\begin{abstract} 
We compare Stetson's photometric standards with measurements listed in a standard 
star catalog constructed using repeated SDSS imaging observations. The SDSS
catalog includes over 700,000 candidate standard stars from the equatorial stripe 
82 ($|$Dec$|<$ 1.266 deg) in the RA range 20h 34' to 4h 00', and with the $r$ band
magnitudes in the range 14--21. The distributions of measurements for individual 
sources demonstrate that the SDSS photometric pipeline correctly estimates random 
photometric errors, which are below 0.01 mag for stars brighter than (19.5, 20.5, 
20.5, 20, 18.5) in {\it ugriz}, respectively (about twice as good as for individual 
SDSS runs). We derive mean photometric transformations between the SDSS {\it gri}
and the {\it BVRI} system using 1165 Stetson stars found in the equatorial stripe 82,
and then study the spatial variation of the difference in zeropoints between the
two catalogs. Using third order polynomials to describe the color terms, we find
that photometric measurements for main-sequence stars can be transformed between
the two systems with systematic errors smaller than a few millimagnitudes. 
The spatial variation of photometric zeropoints in the two catalogs typically 
does not exceed 0.01 magnitude. Consequently, the SDSS Standard Star Catalog for 
Stripe 82 can be used to calibrate new data in both the SDSS {\it ugriz}
and the {\it BVRI} systems with a similar accuracy. 
\end{abstract}

\section{                          Introduction                     }

Astronomical photometric data are usually calibrated using sets of standard stars 
whose brightness is known from previous work. The most notable modern optical 
standard star catalogs are Landolt standards (Landolt 1992) and Stetson standards
(Stetson 2000, 2005). Both are reported on the Johnson-Kron-Cousins system (Landolt
1983 and references therein). 
The Landolt catalog provides magnitudes accurate to 1-2\% in the $UBVRI$ bands for $\sim$500
stars in the $V$ magnitude range 11.5--16. Stetson has extended Landolt's work to 
fainter magnitudes, and provided the community with $\sim$1-2\% accurate magnitudes 
in the $BVRI$ bands for $\sim$15,000 stars in the magnitude range $V\la20$. Most 
stars from both sets are distributed along the Celestial Equator, which facilitates 
their use from both hemispheres.

The data obtained by the Sloan Digital Sky Survey (SDSS, York et al. 2000) can be 
used to extend the work by Landolt and Stetson to even fainter levels, and to increase 
the number of standard stars to almost a million. In addition, SDSS has designed its own 
photometric system ($ugriz$, Fukugita et al. 1996) which is now in use at a large 
number of observatories worldwide. This widespread use of the $ugriz$ photometric 
system motivates the construction of a large standard star catalog with $\sim$1\% 
accuracy. As a part of its imaging survey, SDSS has obtained many scans in the 
so-called Stripe 82 region, defined by $|$Dec$|<$ 1.266 deg and RA approximately 
in the range 20h -- 4h. These repeated observations can be averaged to produce more 
accurate photometry than the nominal 2\% single-scan accuracy (Ivezi\'{c} et al. 
2004). 

We briefly describe the construction and testing of a standard star catalog 
in \S 2, and discuss its comparison with Stetson's photometric standards in \S 3.

\section{          The SDSS standard star catalog for stripe 82 }

\subsection{             Overview of SDSS imaging data                }

SDSS is using a dedicated 2.5m telescope (Gunn et al. 2006) to provide homogeneous 
and deep ($r < 22.5$) photometry in five pass-bands 
(Fukugita et al.~1996; Gunn et al.~1998; Smith 
et al.~2002; Hogg et al. 2002) repeatable to 0.02 mag (root-mean-square scatter,
hereafter rms, for sources not limited by photon statistics, Ivezi\'{c} et al.~2003) 
and with a zeropoint uncertainty of $\sim$0.02-0.03 (Ivezi\'{c} et al.~2004). 
The survey sky coverage of close to $\sim$10,000 deg$^2$ 
in the Northern Galactic Cap, and $\sim$300 deg$^2$ in the Southern Galactic 
Hemisphere, will result in photometric measurements for well over 100 million stars and 
a similar number of galaxies\footnote{The recent Data Release 5 lists photometric
data for 215 million unique objects observed in 8000 deg$^2$ of sky, please
see http://www.sdss.org/dr5/.}. Astrometric positions are accurate to better than 
0.1 arcsec per coordinate (rms) for sources with $r<20.5^m$ (Pier et al.~2003), and 
the morphological information from the images allows reliable star-galaxy separation 
to $r \sim$ 21.5$^m$ (Lupton et al.~2001, 2003, Scranton et al. 2002). 

Data from the imaging camera (thirty photometric, twelve astrometric, and two focus CCDs, 
Gunn et al. 1998) are collected in drift scan mode. The images that correspond 
to the same sky location in each of the five photometric bandpasses (these five images 
are collected over $\sim$5 minutes, with 54 sec for each exposure) are grouped 
together for processing as a field. A field is defined as a 36 seconds (1361 pixels, 
or 9 arcmin) long and 2048 pixels wide (13 arcmin) stretch of drift-scanning data 
from a single column of CCDs (sometimes called a scanline, for more details please see 
Stoughton et al. 2002, Abazajian et al. 2003, 2004, 2005, Adelman-McCarthy et al. 2006). 
Each of the six scanlines 
(called together a strip) is 13 arcmin wide. The twelve interleaved scanlines (or two strips) 
are called a stripe ($\sim$2.5 deg wide).

\subsection{ The photometric calibration of SDSS imaging data  }

\label{photomCalib}

SDSS 2.5m imaging data are photometrically calibrated using a network of calibration stars 
obtained in 1520 41.5$\times$41.5 arcmin$^2$ transfer fields, called secondary patches.
These patches are positioned throughout the survey area and are calibrated using a primary 
standard star network of 158 stars distributed around the Northern sky (Smith et al.~2002). 
The primary standard star network is tied to an absolute flux system by the single F0 subdwarf 
star BD+17$^\circ$4708, whose absolute fluxes in SDSS filters are taken from Fukugita et al. (1996).
The secondary patches are grouped into sets of four, and are observed by the
Photometric Telescope (hereafter PT) in parallel with observations of the primary standards.  
A set of four spans all 12 scanlines of a survey stripe along the width of the stripe, 
and the sets are spaced along the length of a stripe at roughly 15 degree intervals,
which corresponds to an hour of scanning at the sidereal rate. 

SDSS 2.5m magnitudes are reported on the "natural system" of the 2.5m telescope defined by 
the photon-weighted effective wavelengths of each combination of SDSS filter, CCD response, 
telescope transmission, and atmospheric transmission at a reference airmass of 1.3 as 
measured at APO\footnote{Transmission curves for the SDSS 2.5m photometric system are
available at http://www.sdss.org/dr5/instruments/imager/.}. The magnitudes are referred
to as the $ugriz$ system (which is different from the ``primed'' system, $u'g'r'i'z'$, that 
is defined by the PT\footnote{For subtle effects that led to this distinction,
please see Stoughton et al. (2002) and http://www.sdss.org/dr5/algorithms/fluxcal.html.}).
The reported 
magnitudes\footnote{SDSS uses a modified magnitude system (Lupton, Szalay \& Gunn 1999),
which is virtually identical to the standard astronomical Pogson magnitude system at 
high signal-to-noise ratios relevant here.} are corrected for the atmospheric extinction 
(using simultaneous observations of standard stars by the PT) and thus 
correspond to measurements at the top of the atmosphere\footnote{The same atmospheric 
extinction correction is applied irrespective of the source color; the systematic errors
this introduces are probably less than 1\% for all but objects of the most extreme
colors.} 
(except for the fact that the 
atmosphere has an impact on the wavelength dependence of the photometric system response). 
The magnitudes are reported on the AB system (Oke \& Gunn  1983) defined such 
that an object with a specific flux of $F_\nu$=3631 Jy  has $m=0$ (i.e. an object with 
$F_{\nu}$=const. has an AB magnitude equal to the Johnson $V$ magnitude at all wavelengths).
In summary, given a specific flux of an object {\it at the top} of the atmosphere, $F_\nu(\lambda)$,
the reported SDSS 2.5m magnitude in a given band corresponds to (modulo random and systematic errors, 
which will be discussed later)
\begin{equation}
\label{ABmag}
       m = -2.5\log_{10}\left({F_o \over 3631 \, {\rm Jy}}\right),
\end{equation}
where 
\begin{equation}
      F_o = \int{F_\nu(\lambda) \phi(\lambda) d\lambda}.
\end{equation}
Here, $\phi(\lambda)$ is the normalized system response for the given band,
\begin{equation}
      \phi(\lambda) = {\lambda^{-1} S(\lambda) \over \int{\lambda^{-1} S(\lambda) d\lambda}},
\end{equation}
with the overall atmosphere+system throughput, $S(\lambda)$, available from the website 
given above (for a figure showing $\phi(\lambda)$ for the $ugriz$ system see Smol\v{c}i\'{c} 
et al. 2006). 

The quality of SDSS photometry stands out among available large-area optical sky surveys 
(Ivezi\'{c} et al.~2003, 2004; Sesar et al.~2006). Nevertheless, the achieved accuracy is 
occasionally worse than the nominal 0.02-0.03 mag (root-mean-square scatter for sources not limited 
by photon statistics). Typical causes of substandard photometry include an incorrectly 
modeled PSF (usually due to fast variations of atmospheric seeing, or lack of a sufficient 
number of the isolated bright stars needed for modeling the PSF), unrecognized changes in atmospheric 
transparency, errors in photometric zeropoint calibration, effects of crowded fields 
at low Galactic latitudes, an undersampled PSF in excellent seeing conditions ($\la 0.8$ 
arcsec; the pixel size is 0.4 arcsec), incorrect flatfield, or bias vectors, scattered 
light correction, etc. Such effects can conspire to increase the photometric errors to 
levels as high as 0.05 mag (with a frequency, at that error level, of roughly one field 
per thousand). However, when multiple scans of the same sky region are available, many 
of these errors can be minimized by properly averaging photometric measurements.

\subsection{          The catalog construction and internal tests   }

A detailed description of the catalog construction, including the flatfield 
corrections, and various tests of its photometric quality can be found in Ivezi\'{c} et 
al. (2006). Here we only briefly describe the main catalog properties. 

The catalog is based on 58 SDSS runs from stripe 82 (approximately 20h $<$ RA $<$ 04h
and $|$Dec$|<$ 1.266) obtained in mostly photometric conditions (as 
indicated by the calibration residuals, infrared cloud camera\footnote{For more
details about the camera see http://hoggpt.apo.nmsu.edu/irsc/irsc\_doc/.}, 
and tests performed by {\it runQA} quality assessment pipeline\footnote{For a description
of {\it runQA} pipeline see Ivezi\'{c} et al. 2004.}).
Candidate standard stars from each run are selected by requiring  
\begin{enumerate}
\item that objects are unresolved (classified as STAR by the photometric pipeline)
\item
that they have quoted photometric errors (as computed by the photometric pipeline)
smaller than 0.05 mag in at least one band, and 
\item
that processing flags BRIGHT, SATUR, BLENDED, EDGE are not set\footnote{For
more details about photometric processing flags see Stoughton et al. (2002) and
http://www.sdss.org/dr4/products/catalogs/flags.html.}. 
\end{enumerate}
These requirements select unsaturated sources with sufficiently
high signal-to-noise per single observation to approach the final 
photometric errors of 0.02 mag or smaller.

After positionally matching (within 1 arcsec) all detections of a single
star, various photometric statistics such as mean, median, root-mean-square 
scatter, number of observations, and $\chi^2$ per degree of freedom 
($\chi^2_{pdf}$) are computed in each band. This initial catalog of multi-epoch 
observations includes 924,266 stars with at least 4 observations in each of 
the $g$, $r$ and $i$ bands. The median number of observations per star 
and band is 10, and the total number of photometric measurements is 
$\sim$40 million. The errors for the averaged photometry are below 0.01 
mag at the bright end. These errors are reliably computed by photometric pipeline, 
as indicated by the $\chi^2_{pdf}$ distributions.

Adopted candidate standard stars must have at least 4 observations
in each of the $g$, $r$ and $i$ bands and, 
to avoid variable sources, $\chi^2_{pdf}$ less than 3 in the $gri$ bands.
The latter cut rejects about 20\% of stars.
We also limit the RA range to 20h 34' $<$ RA $<$ 04h 00', which provides a 
simple areal definition (together with $|$Dec$|<$1.266 deg) while excluding only 
a  negligible fraction of stars. These requirements result in a catalog with 
681,262 candidate standard stars. Of those, 638,671 have the random error for the median magnitude 
in the $r$ band smaller than 0.01 mag, and 131,014 stars have the random error 
for the median magnitude smaller than 0.01 mag in all five bands.

The internal photometric consistency of this catalog is tested using
a variety of methods, including the position of the stellar locus in the
multi-dimensional color space, color-redshift relations for luminous red galaxies, 
and a direct comparison with the secondary standard star network. While none
of this methods is without its disadvantages, together they suggest that 
the internal photometric zeropoints are spatially stable at the 1\% level.

\section{An external test of catalog quality based on Stetson's standards}

\begin{figure}[!ht]
\plotfiddle{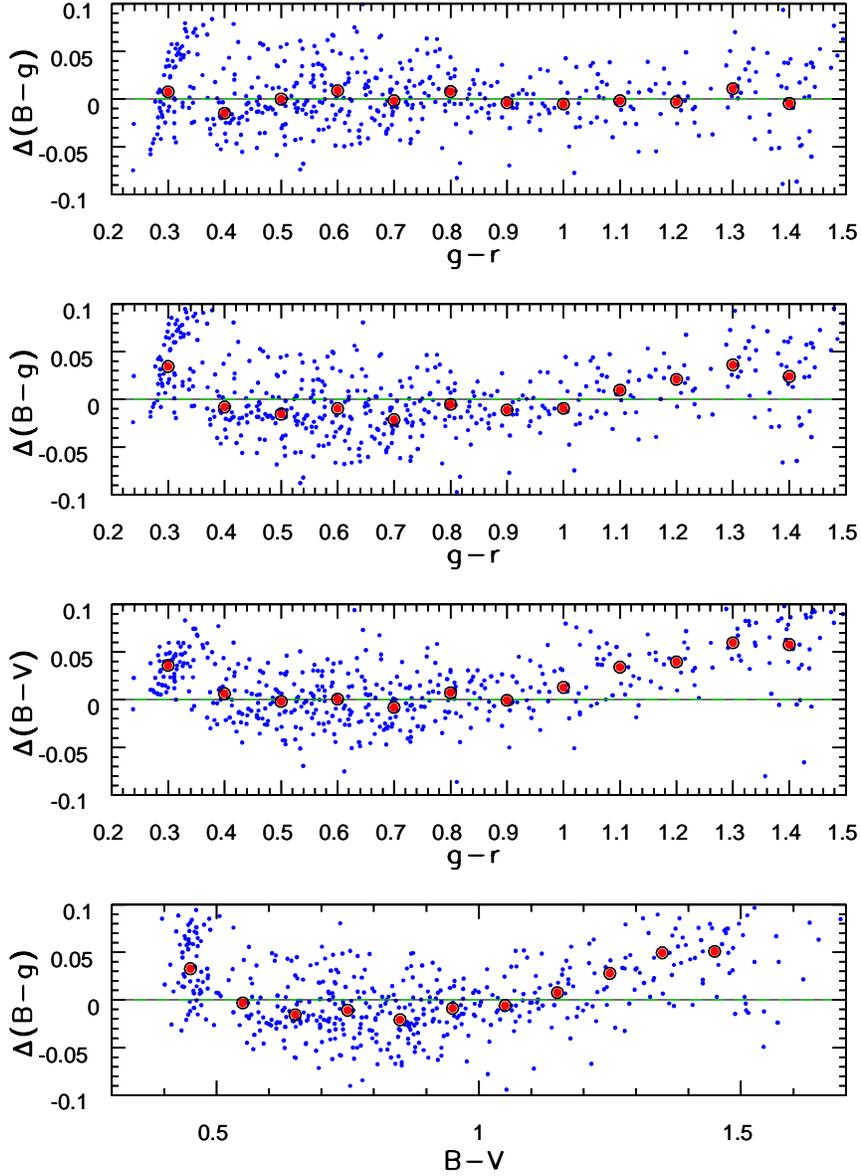}{14.5cm}{0}{75}{75}{-170}{-350}
\caption{An illustration of the need for non-linear color terms when 
transforming SDSS photometry to the BVRI system. 
The small dots in the top panel show the residuals (in magnitudes) for the 
cubic $B-g$ transformation based on eq.~\ref{BVRItrans}, as a function of the 
$g-r$ color. The large 
symbols show the medians for 0.1 mag wide $g-r$ bins. The second panel is 
analogous to the top panel, except 
that a best-fit {\it linear} transformation is used ($B-g=0.345\,(g-r)+0.205$). 
Note the increased deviation of median residuals from zero. The third panel 
shows residuals for the relation $B-V=0.949\,(g-r)+0.197$ and demonstrates that 
even such a color vs. color relation is measurably non-linear. The bottom panel 
shows that this is not a peculiarity of the $g-$ color because a transformation 
based on the $B-V$ color, $B-g=0.364\,(B-V)+0.133$, is also measurably non-linear.
\label{linearity}}
\end{figure}

\begin{table}[!ht]
\caption{SDSS to BVRI transformations}
\smallskip
\begin{center}
{\small
\begin{tabular}{crrrrrrrrr}
\tableline
\noalign{\smallskip}
color & $<>_{med}^a$ & $\sigma_{med}^b$ & $\chi_{med}^c$ & $<>_{all}^d$ & $\sigma_{all}^e$ & A & B & C & D$^f$ \\
\noalign{\smallskip}
\tableline
\noalign{\smallskip}
$B-g$ & -1.6 & 8.7 & 1.4 &  1.0 &  32 &  0.2628 & -0.7952 &  1.0544 &  0.0268  \\
$V-g$ &  0.8 & 3.9 & 1.0 &  0.9 &  18 &  0.0688 & -0.2056 & -0.3838 & -0.0534  \\
$R-r$ & -0.1 & 5.8 & 0.9 &  1.2 &  15 & -0.0107 &  0.0050 & -0.2689 & -0.1540  \\ 
$I-i$ &  0.9 & 6.1 & 1.0 &  1.2 &  19 & -0.0307 &  0.1163 & -0.3341 & -0.3584  \\
\noalign{\smallskip}
\tableline
\end{tabular}
}
\tablenotetext{a}{The median value of median transformation residuals (differences between the measured 
values of colors listed in the first column and those synthesized using eq.~\ref{BVRItrans}) in 0.1 mag 
wide $g-r$ bins for stars with 0.25$< g-r <$1.45 (in millimag). These medians of medians measure the
typical level of systematics in the $gri$-to-$BVRI$ photometric transformations introduced by the adopted
analytic form (see eq.~\ref{BVRItrans}).} 
\tablenotetext{b}{The root-mean-square scatter for median residuals described above (in millimag).} 
\tablenotetext{c}{The root-mean-square scatter for residuals normalized by statistical noise.
The listed values are $\sim$1, which indicates that the scatter around adopted photometric transformations
listed under b) is consistent with expected noise.} 
\tablenotetext{d}{The median value of residuals evaluated for all stars (in millimag).} 
\tablenotetext{e}{The root-mean-square scatter for residuals evaluated for all stars (in millimag).} 
\tablenotetext{f}{Coefficients A--D needed to transform SDSS photometry to the BVRI system (see eq.~\ref{BVRItrans}).} 
\end{center}
\end{table}

While the tests based on SDSS data suggest that the internal photometric zeropoints 
are spatially stable at the 1\% level, it is of course prudent to verify this 
conclusion using an external independent dataset. 
The only large external dataset with sufficient overlap, depth and accuracy to test the 
quality of the Stripe 82 catalog is that provided by Stetson (2000, 2005). 
Stetson's catalog lists photometry in the $BVRI$ bands (Stetson's photometry 
is tied to Landolt's standards) for $\sim$1,200 stars 
in common (most have $V<19.5$). We synthesize the $BVRI$ 
photometry from SDSS $gri$ measurements using photometric transformations 
of the following form
\begin{equation}
\label{BVRItrans}
  m_{\rm Stetson} - \mu_{\rm SDSS} = A\,c^3 + B\,c^2 + C\,c + D,
\end{equation}
where $m=(BVRI)$ and $\mu=(g,g,r,i)$, respectively, and the color $c$ is  
measured by SDSS ($g-r$ for the $B$ and $V$ transformations, and $r-i$ for 
the $R$ and $I$ transformations). The measurements are {\it not} corrected for 
the ISM reddening. 
Traditionally, such transformations are assumed to be linear in 
color\footnote{For various photometric transformations between the SDSS
and other systems, see Abazajian et al. (2005) and 
http://www.sdss.org/dr4/algorithms/sdssUBVRITransform.html.}
We use higher-order terms in eq.~\ref{BVRItrans} because at the 1-2\% level 
there are easily detectable deviations from linearity for all color choices, 
as shown in Fig.~\ref{linearity}. 
The best-fit coefficients for the transformation of SDSS $gri$ measurements 
to the $BVRI$ system\footnote{
The same transformations can be readily used to transform measurements
in the $BVRI$ system to the corresponding $gri$ values because
$B-V=f(g-r)$ and $R-I=f(r-i)$ are monotonous functions).}
are listed in Table~1, as well as low-order statistics for the 
$m_{\rm Stetson} - \mu_{\rm SDSS}$ difference distribution. 
We find no trends as a function of magnitude at the $<0.005$ mag level.

We have also tested for the effects of interstellar dust reddening and 
metallicity on the adopted photometric relations. For about half of stars
in common, the SFD map (Schlegel, Finkbeiner \& Davis 1998) lists $E(B-V)>0.15$.
The differences in median residuals for these stars and those with smaller
$E(B-V)$ (the median $E(B-V)$ are 0.31 and 0.04) are always less than 0.01 mag
(the largest difference is 8 millimag for the $B-g$ transformation). 

\begin{figure}[!ht]
\plotfiddle{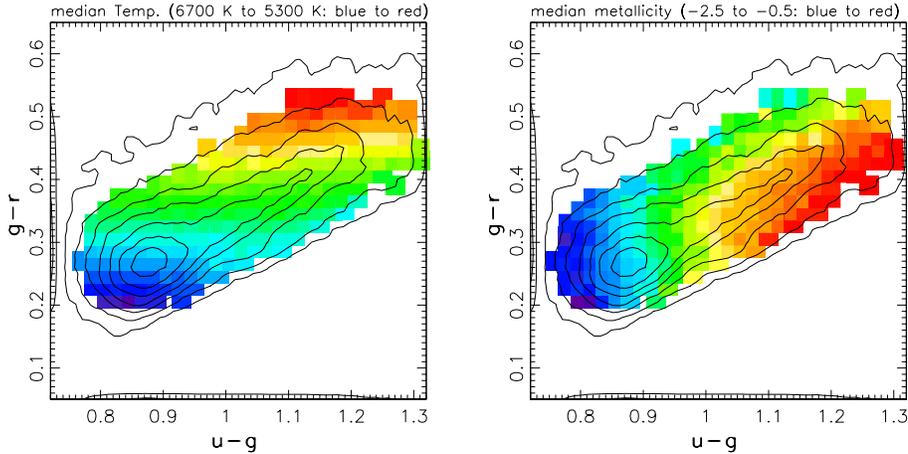}{5.5cm}{0}{90}{90}{-225}{-820}
\caption{
The median effective temperature and median metallicity estimated from SDSS
spectra of $\sim$40,000 stars by Allende Prieto et al. (2006), shown
as a function of the position in the $g-r$ vs. $u-g$ diagram based on SDSS imaging data. 
In the left panel, the temperature in each color-color bin is linearly color-coded 
from 5300 K (red) to 6,700 K (blue). The right panel is analogous, except that 
it shows the median metallicity, linearly color-coded from -0.5 (red) to -2.5 (blue).
\label{specCoding}}
\end{figure}

\begin{figure}[!ht]
\plotfiddle{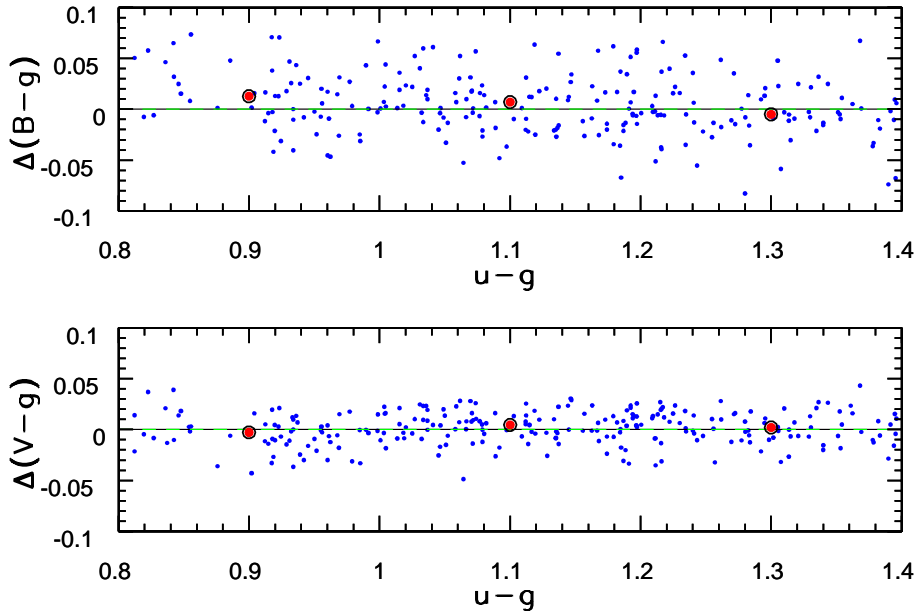}{7.5cm}{0}{80}{80}{-185}{-490}
\caption{An illustration of the effects of metallicity on photometric
transformations. Here the $u-g$ color serves as a proxy for metallicity
(see the right panel in Fig.~\ref{specCoding}). The panels
shows the dependence of the $B-g$ residuals (top) and the $V-g$ residuals 
(bottom) on the $u-g$ color. The effects of metallicity on photometric 
transformations from the SDSS to the BVRI system appear smaller than 
about 0.01 mag, and possibly somewhat larger for the $B-g$ transformation for
stars with $u-g<0.9$. 
\label{metallicity}}
\end{figure}

Stars at the blue tip of the stellar locus with $u-g<1$ are predominantly 
low-metallicity stars (Bond et al. 2006, in prep.), as illustrated in 
Fig.~\ref{specCoding}. Fig.~\ref{metallicity} shows that the median residuals 
for $m_{\rm Stetson} - \mu_{\rm SDSS}$ are the same for the $0.8<u-g<0.95$ 
and $1.0<u-g<1.15$ subsamples to within their measurement errors 
($\sim$10 millimag). There is a possibilty that the offset is somewhat larger 
for the $B-g$ transformation for stars with $u-g<0.9$, but its statistical 
significance is low. If this is a true  effect, it implies a gradient with respect 
to metallicity of about 0.02 mag/dex. 

We conclude that the SDSS catalog described here could also be used to 
calibrate the data to the BVRI system without a loss of accuracy due to 
transformations between the two systems.

\subsection{     The spatial variations of the zeropoints   }

The $BVRI$ photometry from Stetson and that synthesized 
from SDSS agree at the level of 0.02 mag (rms scatter for the magnitude 
differences  of {\it individual} stars; note that the systems are tied to 
each other to within a few millimags by transformations listed in Table~1). 
This scatter is consistent with the claimed accuracy
of both catalogs (the magnitude differences normalized by the 
implied error bars are well described by Gaussians with widths in
the range 0.7--0.8). This small scatter allows us to test for the 
spatial variation of zeropoints between the two datasets, despite
the relatively small number of stars in common.

Stars in common are found in four isolated regions
that coincide with historical and well-known
Kapteyn Selected Areas 113, 92, 95, and 113. 
We determine the zeropoint offsets between the SDSS 
and Stetson's photometry for each region separately
by synthesizing $BVRI$ magnitudes from SDSS $gri$ photometry,
and comparing them to Stetson's measurements.
The implied zeropoint errors (which, of course, 
can be due to either SDSS or Stetson's dataset, or both) are listed
in Table~2. For regions 1-3 the implied errors are only a few
millimags (except for the $B-g$ color in region 1). The discrepancies
are much larger for the three red colors in region 4. A comparison
with the results of internal SDSS tests described by Ivezi\'{c} et al.
(2006) suggests that these discrepancies 
are more likely due to zeropoint offsets in Stetson's photometry for 
this particular region, than to problems with SDSS photometry. 
We contacted P. Stetson who confirmed that his 
observing logs were consistent with this conclusion. Only a small 
fraction of stars from Stetson's list are found in this region. 

Given the results presented in this Section, we conclude\footnote{Here we 
assumed that it is a priori unlikely that the SDSS and Stetson's zeropoint
errors are spatially correlated.} that the rms for the 
spatial variation of zeropoints in the SDSS Stripe 82 
catalog is below 0.01 mag in the $gri$ bands.

\begin{table}[!ht]
\caption{Photometric zeropoint spatial variations}
\smallskip
\begin{center}
{\small
\begin{tabular}{crrrrrrrrrrrr}
\tableline
\noalign{\smallskip}
color & $x_{R1}^a$ & $\sigma_{R1}^b$ & N$_{R1}^c$ & $x_{R2}^a$ & $\sigma_{R2}^b$ & N$_{R2}^c$ & $x_{R3}^a$ & $\sigma_{R3}^b$ & N$_{R3}^c$ & $x_{R4}^a$ & $\sigma_{R4}^b$ & N$_{R4}^c$ \\
\noalign{\smallskip}
\tableline
\noalign{\smallskip}
 $B-g$ &     $-$29 &  21 &   92 &      6 &  27 &  165 &      8 &  42 &  155 &   $-$4 &  27 &  281 \\
 $V-g$ &       0   &  17 &   99 &      0 &  15 &  217 &      6 &  25 &  161 &     17 &  19 &  282 \\
 $R-r$ &     $-$6  &  16 &   58 &      4 &  16 &  135 &   $-$8 &  12 &   11 &     39 &  27 &   60 \\
 $I-i$ &     $-$11 &  16 &   94 &      6 &  18 &  205 &      2 &  16 &  124 &     19 &  15 &   47 \\
\tableline
\end{tabular}
}
\tablenotetext{a}{The median value of residuals (in millimagnitudes) for transformations listed in the first 
column, evaluated  separately for regions 1-4, defined as: 
{\bf R1:} RA$\sim$325, Dec$<$0; {\bf R2:} RA$\sim$15; {\bf R3:} RA$\sim$55;
 {\bf R4:} RA$\sim$325, Dec$>$0.} 
\tablenotetext{b}{The root-mean-square scatter for the transformation residuals (in millimagnitudes).} 
\tablenotetext{c}{The number of stars in each region with good photometry in the required bands.} 
\end{center}
\end{table}

\section{               Discussion and Conclusions                             }

Using repeated SDSS measurements, we have constructed a catalog of 
about 700,000 candidate standard stars. Several independent tests suggest
that both random photometric errors and internal systematic errors in 
photometric zeropoints are below 0.01 mag (about 2-3 times as good as 
individual SDSS runs) for stars brighter than (19.5, 20.5, 20.5, 20, 18.5) 
in {\it ugriz}, respectively. This is by far the largest existing catalog with 
multi-band optical photometry accurate to $\sim$1\%. The catalog is 
publicly available from the SDSS Web Site. 

In this contribution, we have tested the photometric quality of this catalog 
by comparing it to Stetson's standard stars. Using third order polynomials 
to describe the color terms between the SDSS and BVRI systems, we find that 
photometric measurements for main-sequence stars can be transformed between 
the two systems with systematic errors smaller than a few millimagnitudes. 
The spatial variation of photometric zeropoints in the two catalogs typically 
does not exceed 0.01 magnitude. Consequently, the SDSS Standard Star Catalog 
for Stripe 82 can be used to calibrate new data in both the SDSS {\it ugriz}
and the {\it BVRI} systems with a similar accuracy.

\acknowledgements 
Funding for the SDSS and SDSS-II has been provided by the Alfred P. Sloan Foundation, 
the Participating Institutions, the National Science Foundation, the U.S. Department 
of Energy, the National Aeronautics and Space Administration, the Japanese Monbukagakusho, 
the Max Planck Society, and the Higher Education Funding Council for England. The SDSS 
Web Site is http://www.sdss.org/.

The SDSS is managed by the Astrophysical Research Consortium for the Participating 
Institutions. The Participating Institutions are the American Museum of Natural History, 
Astrophysical Institute Potsdam, University of Basel, Cambridge University, Case Western 
Reserve University, University of Chicago, Drexel University, Fermilab, the Institute 
for Advanced Study, the Japan Participation Group, Johns Hopkins University, the Joint 
Institute for Nuclear Astrophysics, the Kavli Institute for Particle Astrophysics and 
Cosmology, the Korean Scientist Group, the Chinese Academy of Sciences (LAMOST), Los 
Alamos National Laboratory, the Max-Planck-Institute for Astronomy (MPIA), the 
Max-Planck-Institute for Astrophysics (MPA), New Mexico State University, Ohio State 
University, University of Pittsburgh, University of Portsmouth, Princeton University, 
the United States Naval Observatory, and the University of Washington.

\bibliographystyle{aa}

\begin{thebibliography}{}
\bibitem[]{DR1} Abazajian, K., Adelman, J.K., Agueros, M.,  et al. 2003, AJ, 126, 2081
\bibitem[]{DR2} Abazajian, K., Adelman, J.K., Agueros, M.,  et al. 2004, AJ, 128, 502
\bibitem[]{DR3} Abazajian, K., Adelman-McCarthy, J.K., Ag\"{u}eros, M.,  et al. 2005, AJ, 129, 1755
\bibitem[Adelman-McCarthy et~al.(2006)]{DR4} Adelman-McCarthy, J.K., et~al. 2006, ApJS, 162, 38
\bibitem[Allende Prieto et al. (2006)]{AP06} Allende Prieto, C., Beers, T.C., Wilhelm, R. et al. 2006, ApJ, 636, 804
\bibitem[]{} Fukugita, M., Ichikawa, T., Gunn, J.E., Doi, M., Shimasaku, K., \& Schneider, 
             D.P. 1996, AJ, 111, 1748
\bibitem[Gunn et al. 1998]{G98} Gunn, J.E., Carr, M., Rockosi, C., et al. 1998, AJ, 116, 3040
\bibitem[Gunn et al. 2006]{G06} Gunn, J.E., Siegmund, W.A., Mannery, E.J., et al. 2006, AJ, 131, 2332
\bibitem[]{} Hogg, D.W., Finkbeiner, D.P., Schlegel, D.J. \& Gunn, J.E. 2002, AJ, 122, 2129
\bibitem[Ivezi\'{c} et al. 2003]{I03} Ivezi\'c, \v Z., Lupton, R.H., Anderson, S., et al.
         2003, Proceedings of the Workshop {\it Variability with Wide Field Imagers}, 
         Mem. Soc. Ast. It., 74, 978 (also astro-ph/0301400)
\bibitem[Ivezi\'{c} et al. (2004)]{AN04} Ivezi\'{c}, \v{Z.}, Lupton, R.H., Schlegel, D.J. et al. 
           2004, AN, 325, 583 
\bibitem[Ivezi\'{c} et al. (2006)]{I06} Ivezi\'{c}, \v{Z.}, Smith, J.A., Miknaitis, G. et al. 
             2006, submitted to AJ
\bibitem[]{} Landolt, A.U. 1983, AJ, 88, 439
\bibitem[]{} Landolt, A.U. 1992, AJ, 104, 340
\bibitem[Lupton, Gunn \& Szalay 1999]{LGS99} Lupton, R.H., Gunn, J.E., \& Szalay, A. 1999, \aj, 118, 1406
\bibitem[Lupton et al. 2001]{Lupton01}Lupton, R.H., Gunn, J.E., Ivezi\'{c}, \v{Z}., Knapp, G.R., 
         Kent, S. \& Yasuda, N. 2001, in ASP Conf. Ser. 238, Astronomical Data Analysis Software 
         and Systems X, eds. F. R. Harnden, Jr., F. A. Primini, and H. E. Payne (San Francisco: Astr. 
         Soc. Pac.), p. 269 (also astro-ph/0101420)
\bibitem[Lupton et al. 2003]{Lupton04} Lupton, R.H., Ivezi\'{c}, \v{Z}., Gunn, J.E., Knapp, G.R., 
         Strauss, M.A. \& Yasuda, N. 2003, in ``Survey and Other Telescope Technologies and 
         Discoveries'', Tyson, J.A. \& Wolff, S., eds. Proceedings of the SPIE, 4836, 350 
\bibitem[Oke \& Gunn 1983]{OG83}Oke, J.B., \& Gunn, J.E. 1983, ApJ, 266, 713
\bibitem[Pier et al. (2003)]{Pier03}Pier, J.R., Munn, J.A., Hindsley, R.B., Hennesy, G.S., Kent,
         S.M., Lupton, R.H. \& Ivezi\'{c}, \v{Z}. 2003, AJ, 125, 1559
\bibitem[Schlegel, Finkbeiner \& Davis 1998]{SFD98}Schlegel, D., Finkbeiner,D.P.
                   \& Davis, M. 1998, \apj 500, 525
\bibitem[]{}Scranton, R., Johnston, D., Dodelson, S., et al. 2002, ApJ, 579, 48
\bibitem[Sesar et al. 2006]{} Sesar, B., Svilkovi\'{c}, D., Ivezi\'{c}, \v{Z.}, et al. 2006,
               AJ, 131, 2801
\bibitem[]{} Smith, J.A., Tucker, D.L., Kent, S.M., et al. 2002, AJ, 123, 2121
\bibitem[]{} Smol\v{c}i\'{c}, V., Ivezi\'{c}, \v{Z}., Ga\'{c}e\v{s}a, M., et al. 2006, MNRAS, 371, 121
\bibitem[]{} Stetson, P.B. 2000, PASP, 112, 925
\bibitem[]{} Stetson, P.B. 2005, PASP, 117, 563
\bibitem[Stoughton et al. (2002)]{EDR} Stoughton, C., Lupton, R.H., Bernardi, M., et al. 2002, AJ, 123, 485
\bibitem[York et al. (2000)]{SDSS} York, D.G., Adelman, J., Anderson, S., et al. 2000, AJ, 120, 1579 
\end{thebibliography}
{}

\end{document}